# Revisiting Interactions of Multiple Driver States in Heterogenous Population and Cognitive Tasks

Jiyao Wang, *Student Member, IEEE*, Ange Wang, Song Yan, Dengbo He, and Kaishun Wu, *Fellow, IEEE*

***Abstract*— In real-world driving scenarios, multiple states occur simultaneously due to individual differences and environmental factors, complicating the analysis and estimation of driver states. Previous studies, limited by experimental design and analytical methods, may not be able to disentangle the relationships among multiple driver states and environmental factors. This paper introduces the Double Machine Learning (DML) analysis method to the field of driver state analysis to tackle this challenge. To train and test the DML model, a driving simulator experiment with 42 participants was conducted. All participants drove SAE level-3 vehicles and conducted three types of cognitive tasks in a 3-hour driving experiment. Drivers' subjective cognitive load and drowsiness levels were collected throughout the experiment. Then, we isolated individual and environmental factors affecting driver state variations and the factors affecting drivers' physiological and eye-tracking metrics when they are under specific states. The results show that our approach successfully decoupled and inferred the complex causal relationships between multiple types of drowsiness and cognitive load. Additionally, we identified key physiological and eye-tracking indicators in the presence of multiple driver states and under the influence of a single state, excluding the influence of other driver states, environmental factors, and individual characteristics. Our causal inference analytical framework can offer new insights for subsequent analysis of drivers' states. Further, the updated causal relation graph based on the DML analysis can provide theoretical bases for driver state monitoring based on physiological and eye-tracking measures.**

## I. INTRODUCTION

Driver drowsiness and high cognitive load can both negatively impact driving safety. Specifically, drowsiness, characterized by decreased alertness, involves diminished executive functioning, mental effort, and involuntary muscle inhibition [1]. Each year, it is estimated that drowsy driving contributes to around 328,000 crashes in the US, including 109,000 injury-related crashes and 6,400 fatal crashes [2]. At the same time, the cognitive workload is defined as the information processing capacity or cognitive resources needed to complete a task [3]. While autonomous technologies promise to reduce drivers' cognitive workload and drowsiness by relieving drivers from driving tasks [4], before fully autonomous vehicle comes, human drivers still must share control with driving automation systems. Particularly, with the Society of Automotive Engineers (SAE) Level-3 (L3) advanced driving systems (ADS) [5], the vehicle can control both steering and acceleration/deceleration but still requires the driver to remain actively engaged and ready to take over at any moment. Given that drivers are inclined to engage in non-driving-related tasks (NDRTs) with the assistance of driving automation [6], understanding the impact of NDRTs on the drivers' states is essential to the driving safety of SAE Level-3 vehicles.

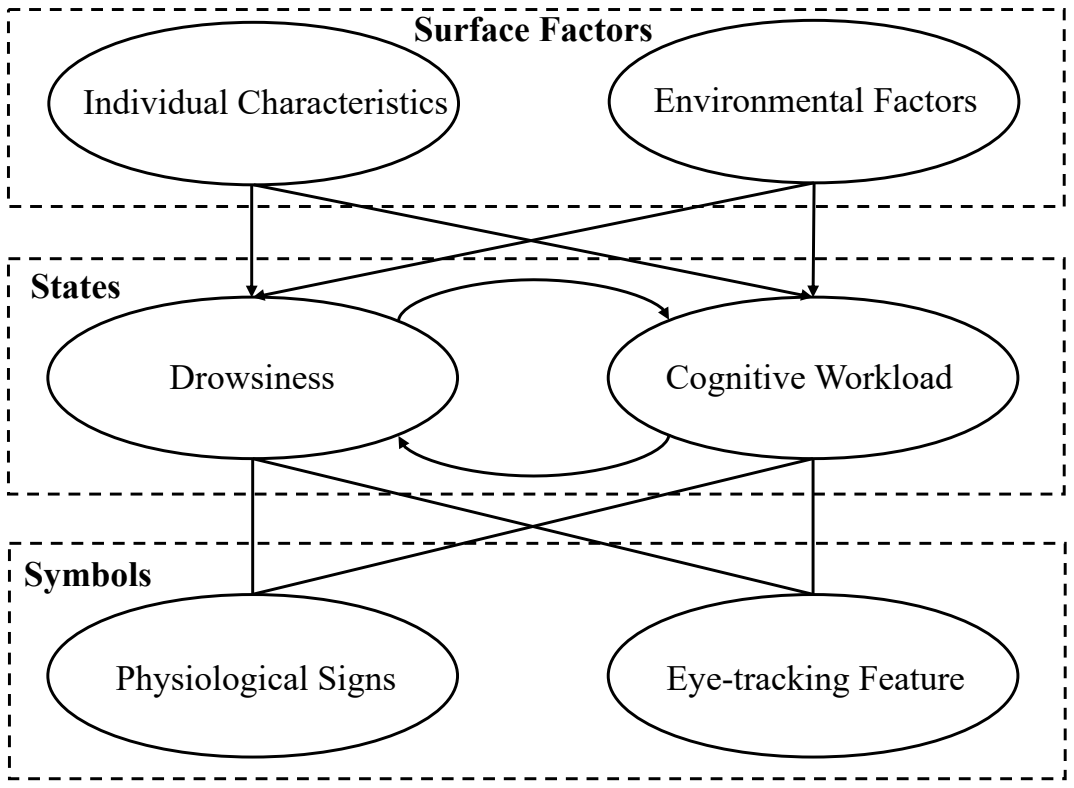

**Fig. 1.** The causal relation graph from surface factors and individual factors to the symbols caused by driver states. Note that, as the symbols are the representation of states, there are no causalities but correlations between symbols and states.

Since the cognitive resource is multi-dimensional [7], different dimensions of NDRT tasks can bring disparities in the cognitive load states of drivers. For example, it has also been suggested that the progress of sole driving tasks can be accompanied by insufficient cognitive load, leading to drowsiness and impairing driving safety [8], [9], a moderate level of cognitive load due to NDRTs may reduce drowsiness levels [10], but a too-high cognitive load can cause delayed reactions in emergencies [11], a limited field of vision [12], and a reduced ability to foresee potential hazards [13]. However, during a drive, the effect of long-time driving and the high cognitive load as a result of NDRTs may co-exist, leading to

This research is supported by the start-up funding of the Hong Kong University of Science and Technology (Guangzhou). This work is supported by This work was supported by the Natural Science of China Foundation of Guangdong Province (2024A1515010392), and partially by the National Natural Science Foundation of China (grant no. 52202425), the Project of Hetao Shenzhen-Hong Kong Science and TechnologyInnovation Cooperation Zone (HZQB-KCZYB-2020083).

Jiyao Wang, Ange Wang, Song Yan, and Dengbo He is with the Systems Hub, the Hong Kong University of Science and Technology (Guangzhou), China, (e-mail: jwanggo@connect.ust.hk; awang324@connect.hkust-gz.edu.cn; syan931@connect.hkust-gz.edu.cn; dengbohe@hkust-gz.edu.cn). Kaishun Wu is with the Information Hub, the Hong Kong University of Science and Technology (Guangzhou), China, (e-mail: wuks@hkust-gz.edu.cn).

compounded effects on drivers' readiness and driving safety.

Understanding the factors leading to specific driver states and the associated symbols may guide the countermeasures alleviating the negative effects of specific states. However, though previous research tried to understand the relationships between NDRTs and specific driver states with specific variables controlled in experiments, isolating the effects of specific factors on a state is challenging. Specifically, statistical regression analysis is a classical method and has been widely used in previous works [6], [8], but linear regression cannot eliminate the effect of uncontrolled confounders on the dependent variables. For example, with the progress of an experiment, the subjects performing the cognitive load induction task (typically NDRTs) may have developed both drowsiness as a result of long experiment duration and cognitive load as a result of the NDRT. Thus, conclusions retrieved from previous studies may have been biased by these compound effects and whether specific NDRT affects cognitive load directly or through driving fatigue is unknown.

In addition, past research has pointed out that individual heterogeneity [14] has a significant influence on a driver's cognitive state [15], [16] and fatigue development [17] during driving. However, it is still difficult to tell whether one demographic feature impacts cognitive load and drowsiness directly or through affecting other states of drivers. Similarly, when measuring fatigue and cognitive workload through physiological and eye-tracking measures [18], [19], linear correlation analyses may be insufficient to decouple the dual effect of the states. Hence, new approaches to isolate the co-effects of driver states are needed.

Inspired by the research in economics, a double/debiased machine learning (DML) [20] approach is introduced in our work. By specifying the confounding variable **W**, feature **X**, the treatment variable **T**, and outcome **Y**, the DML is able to precisely pinpoint the direct causal effect of concern and eliminate the spurious bias from confounders based on Neyman-orthogonal and K-fold cross-fitting [21]. According to the prior knowledge from previous research, we propose a causal relation graph in Figure 1. Specifically, the influence of individual heterogeneity from the upper layer can affect the States and Symbols in the lower layer and we propose four research questions (RQs):

**RQ1**: How are driver states affected by NDRTs and the progress of the experiment (i.e., time) jointly? This RQ simulates the realistic scenarios where the NDRTs and driving duration are entangled.

**RQ2**: Excluding the influence of NDRT and time, whether and how does individual heterogeneity affect cognitive load and drowsiness? This RQ explores the influence of individual heterogeneity on drivers' states, as has been studied but still may suffer from bias in previous research [15], [16].

**RQ3**: Excluding the influence of individual heterogeneity, what are the compound effects of driver states on symbols? This RQ may answer in a naturalistic setting, how the symbols associated with multiple driver states, excluding the influence of individual differences, given that multiple states may co-exist in drivers.

**RQ4**: Excluding the influence of individual heterogeneity and cross-effect between states, what are the relationships between symbols and a single state? This RQ can help clarify the relationships between a specific driver state and driver symbols, guiding more targeted driver state monitoring.

To answer the above questions, a driving simulator experiment was conducted to assess the impact of high cognitive load on drivers in SAE Level-3 vehicles. Three standardized cognitive tasks were used to impose different types of cognitive load: the n-back tasks [22], which require memory resources; mathematical tasks [23], which require numerical processing resources; and spatial reasoning tasks [24], which require spatial resources. The conclusions obtained in this paper can support the clarification of the complex relationship between driver states and state-related measures in driving conditions and guide the optimization of driver monitoring systems.

## II. Related Works

### *A. Impact of Individual Characteristics and Environmental Factors on Driver States*

Drowsiness and cognitive workload are critical factors influencing driving performance and safety and have been found to be affected by many external factors. For example, older drivers often experience higher baseline cognitive load and mental fatigue [25], [26] due to lower upper limits of cognitive resources [27], [28]. Other research reported that gender [9] and driving experiences [29] could affect distribution of driver states. In addition to demographic features, psychological factors [30] were also found to be correlated with driver states. For instance, trust can affect drivers' information processing abilities [31] , and overtrust in autonomous driving can lead to drowsiness [32].

In addition, external factors, such as NDRTs can also induce high cognitive loads and impair the takeover quality in safety-critical events [33]. Given the correlation between drowsiness and cognitive load [1], NDRTs were also found to reduce drowsiness in driving [9]. However, given the diversity of NDRTs, the different effects of NDRTs with different modalities should be further discussed.

### *B. Physiological and Eye-Tracking Measure of Driver States*

Numerous studies have explored how cognitive load impacts eye-tracking measures. For instance, on-road studies by Recarte and Nunes [12], [34] demonstrated that performing secondary cognitive tasks while driving can increase pupil diameter. Similarly, [35] found that increased visual demand can enlarge pupil diameter; while physical resource demands can increase heart rate. Chen et al. also [15] suggested that changes in pupil diameter, saccade frequency and duration, fixation length, and 3D gaze entropy are reliable indicators of a driver's cognitive load in semi-autonomous driving scenarios. Besides eye-tracking indicators, physiological measures have been extensively used to assess the impact of driver state. In [36], it was found that engaging in NDRTs can result in elevated heart rate (HR) [37] and reduced heart rate variability (HRV) [36]. Additionally, Skin Conductance Response (SCR) and Respiratory Rate (RR) are also found [37] to be strongly associated with cognitive task difficulty [22].

As for drowsiness, eye-tracking metrics are found to be robust indicators. Schleicher [38] found that the blinks and saccades are highly correlated with drowsiness and thus can be used to predict lapses in attention. Besides, other research found that increased blink duration and frequency are also reliable indicators of drowsiness [39], [40] and can be used for real-time monitoring of a driver's states [41], [42]. In addition, electroencephalography (EEG) and HRV were also used to assess drowsiness in previous research [43], [44].

Although some works tried to investigate the physiological and eye-related indicators under specific drowsiness types (e.g., mental fatigue [45] or physical fatigue [46]), few empirical attempts have been made to decouple the cross-effects of driver states on physiological and oculomotor indicators.

## III. Methodology

### *A. Experiment Design*

#### 1) Experiment Conditions and NDRTs

A driving simulation experiment with a within-subject design was adopted. By varying the types and difficulty levels of cognitive NDRTs, we aim to understand how these different demands can influence drives' physiological and eye movement responses. Table I and Figure 2 present an overview of the three types of cognitive tasks (6 specific tasks) we adopted in this study plus a baseline without NDRT tasks. For each NDRT, each participant went through 3 drives, leading to 21 drives in total. A Latin-square design was adopted to minimize the effect of trail order, leading to 21 orders in total.

#### 2) Driving Task

All drives were on two-way six-lane highways with a speed limit of 120 kilometers per hour and a traffic density of 6 vehicles per kilometer per lane. However, the top speed of the driving automation was 110 kilometers per hour. In manual driving mode, drivers were required to drive in the middle lane. Each drive was approximately 7 kilometers long.

#### 3) Experiment Procedure

Participants were instructed to maintain normal sleep patterns, abstain from alcohol, and avoid caffeine intake for 24 hours prior to the experiment. Upon arrival, written consent was obtained. Subsequently, they underwent a half-hour training session covering the experimental procedure, vehicle operation, cognitive tasks, and subjective questionnaires. Then, their initial trust in driving automation was assessed through a questionnaire by [47]. Finally, the data collection devices, including physiological sensors and eye trackers, were put on and calibrated before they started 21 experimental drives. After each drive, the participant was asked to complete a questionnaire to assess their states in the previous drive. Specifically, cognitive load was measured using the NASA Task Load Index (NASA-TLX) [48] and drowsiness level was measured using the Karolinska Sleepiness Scale (KSS) [49]. The total duration of the experiment for each participant was approximately 3 hours (about 1.5 hours for driving).

TABLE I: Summary of Cognitive Load Tasks (NDRTs) Used in Our Study.

| Task Type | Description | Task Level(s) | Cognitive Resource |
|---|---|---|---|
| N-back Task [22] | A series of stimuli numbers are presented with a pause between each. Participants recall and verbally report the stimulus that is *n* positions earlier. | 0-back (NB0), 1-back (NB1), 2-back (NB2) tasks. | Memory |
| Math Task [23] | Oral backward counting from 3,000 by increments of 3 or 5. | Counting backward by 3 (MT1) or 5 (MT2) from 3,000. | Calculation |
| Spatial Task [24] | Participants listen to an audio clip describing a route and identify the direction faced at the end, simulating cognitive task in navigation tasks. | "What direction is this person when he goes to the north station and moves two stations clockwise?" (Answer: East) (ST) | Spatial Processing |

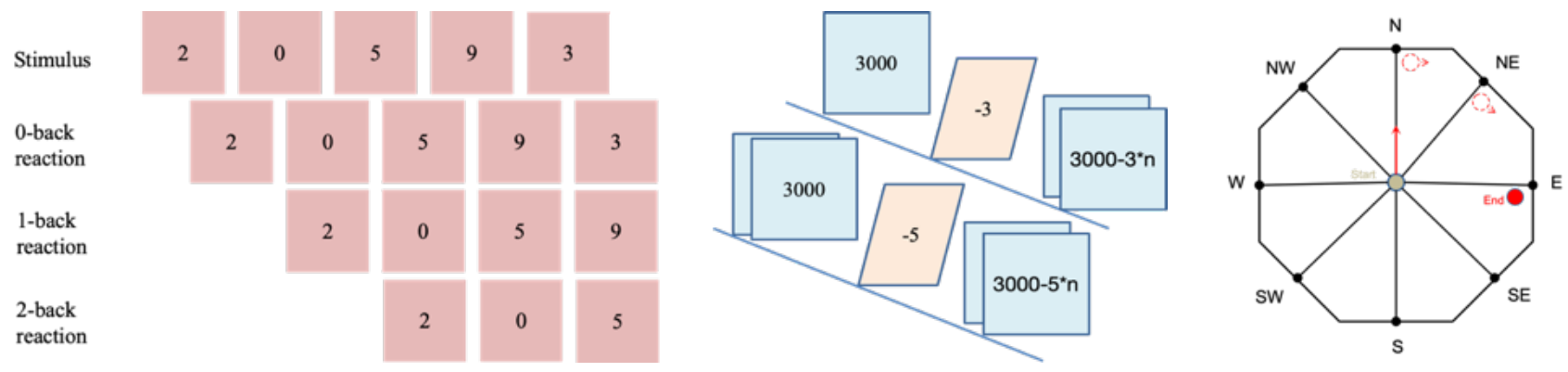

(a) (b) (c)

**Fig. 2.** Cognitive task design, including (a) n-back task, (b)math task, and (c) spatial task.

*B. Participants*

A preliminary sample size calculation was conducted using MorePower [50] before the experiment. The results indicated that a minimum of 24 participants would be required to achieve 80% power at a 95% confidence interval (CI), with an effect size of $\sigma^2 = 0.06$. In this study, a total of 42 drivers (24 males, and 18 females) were recruited. Participants' ages were uniformly distributed from 20 to 60 years, thereby enabling the analysis of age effect. All participants were required to have no prior experience with driving automation and hold a valid driving license for at least one year. Participants were compensated at a rate of 70 RMB per hour. In total, 882 drives were conducted (42 participants and 21 drives per participant). After cleaning the data, due to technical issues, 820 out of the 882 drives were kept for further analysis. This study was approved by the Human and Artefacts Research Ethics Committee at the Hong Kong University of Science and Technology (protocol number: HREP-2023-0199).

*C. Apparatus*

A fixed-base driving simulator was adopted, with three 42-inch screens showing a horizontal view angle of 150° and a vertical viewing angle of 47°, and one external tablet with two touch buttons to engage and disengage the driving automation. The Silab 7.1 by WIVW was used to develop driving scenarios and collect vehicle operation data at 60 Hz. Participant's eye movement data was recorded using a head-mounted eye tracker, the Dikabilis Glass 3 by Ergoneers, and mapped into the front view video with a resolution of 1920*1080, captured by a scene camera between the two eyes at 60 Hz. All physiological data, including Electrocardiogram (ECG), Electrodermal Activity (EDA), and Respiratory (RESP), were collected using the sensors by Ergoneers at 100Hz.

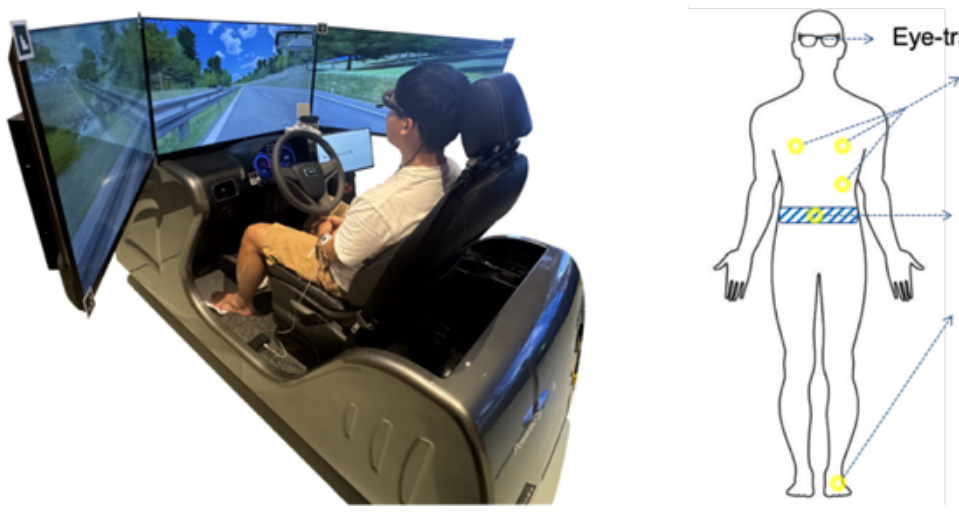

(a) (b)

**Fig. 3.** (a) equipment of the driving platform, and (b) physiological sensors.

*D. Analysis Method*

**1) Variable Extraction**

The cognitive load level and drowsiness level were extracted based on the standard data analysis approach of NASA and NASA [49]. The eye-tracking metrics and physiological metrics were extracted as below:

- Eye-tracking features: These include pupil area (PA, unit: pixel), fixation rate (FR, unit: times/minute), fixation time (FT, unit: s/min), saccade rate (SR, unit: times/minute), saccade time (ST, unit: s/min), and saccade angle (SA, unit: degrees).
- Electrodermal Activity (EDA) Metrics: After applying a fourth-order Butterworth low-pass filter (5Hz) to mitigate high-frequency disturbances, we extracted the skin conductance level (SCL, unit: μS), and skin conductance response (SCR, unit: μS).
- Electrocardiogram (ECG) Metric: ECG signals were processed through a band-pass filter (3 Hz to 45 Hz) [51], followed by R-wave detection using an enhanced Pan-Tompkins algorithm. The extracted metrics include Heart rate (HR, unit: beats/minute), Root Mean Square of Successive Differences between normal heartbeats (RMSSD, unit: ms), Standard deviation of Normal to Normal R Wave (SDNN, unit: ms), Low Frequency (LF, unit: Hz), High Frequency (HF, unit: Hz), and LF /HF.
- Respiratory (RESP) Metrics: After applying a band-pass filter with 0.1 Hz to 0.35 Hz [52], we extracted respiratory rate (RR, unit: respirations/minute), respiratory depth (RD, unit: mm), and respiratory variation (RV, unit: %).

As shown in Figure 1, we considered two types of factors that can affect the driver state and the associated symbols, i.e., individual characteristics, and environmental factors, with the distribution of them presented in Table II. The environmental factors include two variables: NDRT with 7 levels, and a continuous variable, driving time (i.e., 21 drives, each labeled with one distinct number following the experimental order).

**2) Double/Debiased Machine Learning**

DML was first proposed in [20] and widely used in econometrics [53]. It integrates machine learning methods with traditional statistical inference to estimate causal effects. This approach ensures that the estimation of the causal effect remains robust and unbiased. The term "double" comes from the two-step process involved: 1) get residual from the outcome model predicting the relationship between outcome **Y** and features **X** and **W**; 2) get residual from the treatment model which models the effect of features **X** and **W** on the treatment **T**. Then, fit a new model targeting the residual of **Y** using **X** and residual of **T** to obtain unbiased estimation and control impacts of uncontrollable confounders **W**.

Compared to statistical regression, DML also retains the explanatory power of statistical inference. Except for the better performance on high-dimensional data and avoiding over-fitting [20], DML can get the effect of applying **T** (i.e., Average Treatment Effect, ATE) by simulating an experimental-control group on the same batch of samples with **X** feature, and exclude the interference of **W**. Another advantage of DML is that it can obtain the Conditional Average Treatment Effect (CATE) of **T** for **X**, i.e., the difference in the average effect of **T** when it is applied to a specific group of samples [54]. We adopted the "*LinearDML*" function in the 0.15.0 version of "*EconML"* package [55] in Python for modeling. Compared to other DML methods, *LinearDML* in *EconML* adopts the linear parametric method and enables the model to have interpretable model parameters. The two-step process was instantiated by gradient boost machines following previous work [56], and a 5-fold cross-fitting was used to avoid over-fitting. To better present the differences in the driver population due to individual heterogeneity, we used "*SingleTreeCateInterpreter*" function in *EconML* to interpret the model.

In total, 8 models were built to answer the four research questions. A brief illustration of models is provided in Figure 4. Specifically, Model (a)(b) is for RQ1, (c)(d) is for RQ2, (e)(f) is for RQ3, and (g)(h) is for RQ4.

TABLE II: SUMMARY OF THE VARIABLES.

| Variable | Definition & Distribution |
|---|---|
| Age$^{I}$ | The age of participants measured by years of old.<br>- mean: 35.4 (SD: 9.1, min: 23, max: 53) |
| Gender$^{I}$ | 1. Male (n=24, 57.1%)<br>2. Female (n=18, 42.9%) |
| Trust$^{I}$ | The score of trust in ADS.<br>- mean: 36.7 (SD: 6.9, min: 22, max: 49) |
| DriveE$^{I}$ | The years since the participant first obtained the driving license.<br>- mean: 36.7 (SD: 6.9, min: 22, max: 49) |
| DriveD$^{I}$ | The driving distance in past one year.<br>1. <5,000km (n=10, 23.8%)<br>2. 5,000~10,000km (n=12, 28.6%)<br>3. 10,000~20,000km (n=10, 23.8%)<br>4. - >20,000km (n=10, 23.8%) |
| KSS$^{S}$ | The drowsiness level of participants.<br>- mean: 4.2 (SD: 1.9, min: 1, max: 10) |
| NASA$^{S}$ | The cognitive workload level of participants.<br>- mean: 8.6 (SD: 5.2, min: 1, max: 20) |

*Note: The superscript 'I' means this variable belongs to Individual Characteristics, and 'S' indicates it is States.*

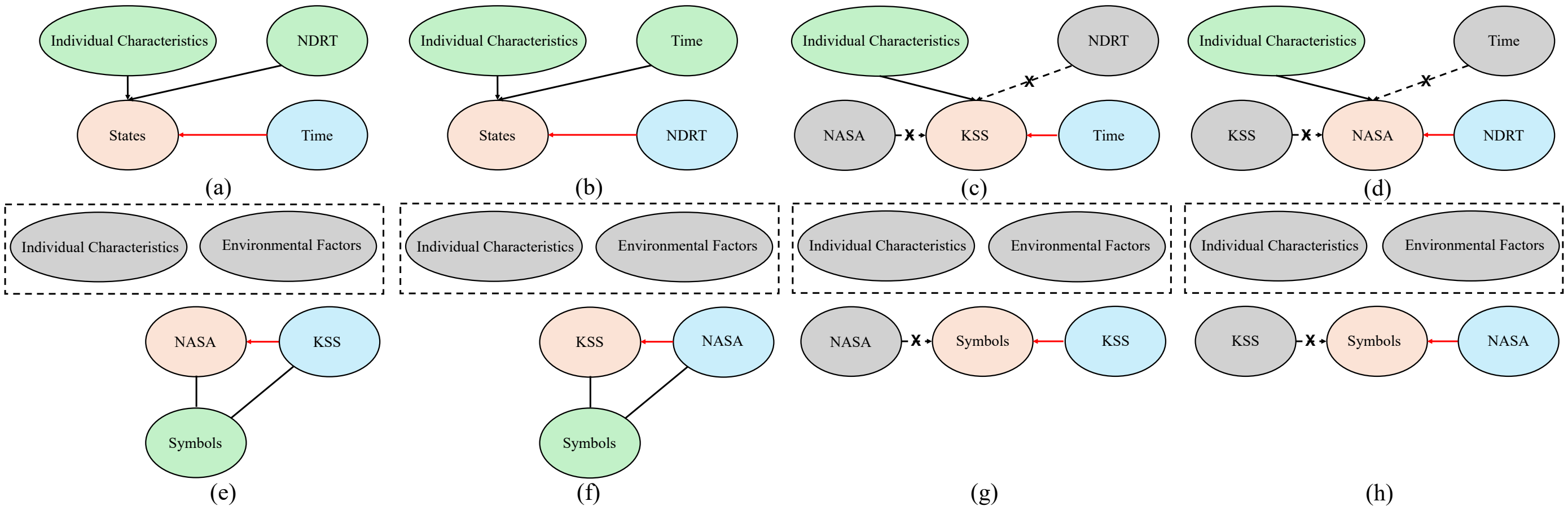

**Fig. 4.** Brief illustration of constructed models. The green node indicates feature **X**; The orange node means outcome **Y**; The grey node means confounders **W**; and the blue node means treatment **T**. We are interested in the coefficients of the model of the black edge, and the treatment effect on the outcome for red edges.

TABLE III: SIGNIFICANT ($P < .05$) COEFFICIENTS OF MODELS.

| Model | X | Y | T | Estimation | SE | Z Stat | *p*-value | 95%CI-lower | 95%CI-upper |
|---|---|---|---|---|---|---|---|---|---|
| **(a)** | Trust | NASA | Time | 0.007 | 0.003 | 2.517 | .01 | 0.002 | 0.011 |
| | DriveD | KSS | Time | 0.019 | 0.009 | 2.084 | .04 | 0.004 | 0.035 |
| **(b)** | Trust | NASA | NB1 | 0.182 | 0.063 | 2.884 | .004 | 0.078 | 0.286 |
| | Age | NASA | ST | 0.315 | 0.078 | 4.021 | <.0001 | 0.186 | 0.444 |
| | Age | KSS | NB1 | 0.064 | 0.026 | 2.509 | .01 | 0.022 | 0.107 |
| | Age | KSS | NB2 | 0.052 | 0.025 | 2.112 | .04 | 0.012 | 0.093 |
| **(d)** | Trust | NASA | NB1 | 0.154 | 0.064 | 2.382 | .02 | 0.048 | 0.260 |
| | Age | NASA | ST | 0.176 | 0.077 | 2.294 | .02 | 0.050 | 0.303 |
| **(e)** | SCR | NASA | KSS | -0.122 | 0.058 | -2.098 | .04 | -0.218 | -0.026 |
| **(f)** | SCR | KSS | NASA | -0.015 | 0.007 | -2.095 | .04 | -0.026 | -0.003 |
| | RMSSD | KSS | NASA | 0.029 | 0.009 | 3.068 | .002 | 0.013 | 0.045 |
| | SDNN | KSS | NASA | -0.053 | 0.017 | -3.073 | .002 | -0.081 | -0.025 |
| | HF | KSS | NASA | 0.169 | 0.079 | 2.134 | .03 | 0.039 | 0.299 |

*Note: In this table, for the effect of discrete treatment NDRT on the association between **X** and **Y**, the baseline is the control group **T0**.*

TABLE IV: SIGNIFICANT ($P < .05$) ATE OF DISCRETE TREATMENT NDRT ANALYSIS OF MODELS.

| Model | Y | T0 | T1 | Estimation | SE | Z Stat | *p*-value | 95%CI-lower | 95%CI-upper |
|---|---|---|---|---|---|---|---|---|---|
| **(b)** | NASA | Base | NB0 | 2.150 | 0.389 | 5.519 | <.0001 | 1.509 | 2.790 |
| | | | NB1 | 4.891 | 0.401 | 12.182 | <.0001 | 4.230 | 5.551 |
| | | | NB2 | 9.342 | 0.427 | 21.870 | <.0001 | 8.639 | 10.044 |
| | | | MT1 | 6.910 | 0.438 | 15.775 | <.0001 | 6.190 | 7.631 |
| | | | MT2 | 4.106 | 0.410 | 10.017 | <.0001 | 3.431 | 4.780 |
| | | | ST | 8.487 | 0.456 | 8.615 | <.0001 | 7.737 | 9.237 |
| | | NB0 | NB1 | 2.741 | 0.386 | 7.093 | <.0001 | 2.105 | 3.377 |
| | | | NB2 | 7.192 | 0.413 | 17.410 | <.0001 | 6.512 | 7.871 |
| | | | MT1 | 4.760 | 0.424 | 11.225 | <.0001 | 4.063 | 5.458 |
| | | | MT2 | 1.956 | 0.395 | 4.946 | <.0001 | 1.305 | 2.606 |
| | | | ST | 6.338 | 0.443 | 14.297 | <.0001 | 5.608 | 7.067 |
| | | NB1 | NB2 | 4.451 | 0.424 | 10.491 | <.0001 | 3.753 | 5.149 |
| | | | MT1 | 2.019 | 0.439 | 4.605 | <.0001 | 1.298 | 2.741 |
| | | | ST | 3.597 | 0.456 | 7.893 | <.0001 | 2.847 | 4.346 |
| | | NB2 | MT1 | -2.431 | 0.460 | -5.288 | <.0001 | -3.188 | -1.675 |
| | | | MT2 | -5.236 | 0.434 | -12.074 | <.0001 | -5.949 | -4.523 |
| | | MT1 | MT2 | -2.804 | 0.446 | -6.288 | <.0001 | -3.538 | -2.071 |
| | | | ST | 1.577 | 0.488 | 3.229 | .001 | 0.774 | 2.381 |
| | | MT2 | ST | 4.382 | 0.463 | 9.456 | <.0001 | 3.620 | 5.144 |
| | KSS | Base | NB0 | -0.334 | 0.140 | -2.380 | .02 | -0.565 | -0.103 |
| | | | NB1 | -0.410 | 0.147 | -2.782 | .005 | -0.653 | -0.168 |
| | | | NB2 | -0.713 | 0.149 | -4.793 | <.0001 | -0.958 | -0.469 |
| | | | MT1 | -0.625 | 0.140 | -4.477 | <.0001 | -0.855 | -0.396 |
| | | | MT2 | -0.344 | 0.146 | -2.358 | .02 | -0.585 | -0.104 |
| | | | ST | -0.710 | 0.133 | -5.354 | <.0001 | -0.928 | -0.492 |
| | | NB0 | NB2 | -0.379 | 0.148 | -2.563 | .01 | -0.622 | -0.136 |
| | | | MT1 | -0.291 | 0.139 | -2.091 | .04 | -0.520 | -0.062 |
| | | | ST | -0.376 | 0.132 | -2.845 | .004 | -0.593 | -0.159 |
| | | NB1 | ST | -0.300 | 0.139 | -2.152 | .03 | -0.529 | -0.071 |
| | | NB2 | MT2 | 0.369 | 0.154 | 2.404 | .02 | 0.117 | 0.622 |
| | | MT2 | ST | -0.366 | 0.138 | -2.652 | .008 | -0.592 | -0.139 |
| **(d)** | NASA | Base | NB0 | 1.416 | 0.394 | 3.597 | <.0001 | 0.768 | 2.063 |
| | | | NB1 | 4.037 | 0.395 | 10.216 | <.0001 | 3.387 | 4.687 |
| | | | NB2 | 8.333 | 0.450 | 18.519 | <.0001 | 7.593 | 9.073 |
| | | | MT1 | 6.238 | 0.438 | 14.236 | <.0001 | 5.517 | 6.958 |
| | | | MT2 | 3.645 | 0.384 | 9.484 | <.0001 | 3.013 | 4.278 |
| | | | ST | 7.310 | 0.445 | 16.418 | <.0001 | 6.578 | 8.043 |
| | | NB0 | NB1 | 2.622 | 0.377 | 6.952 | <.0001 | 2.001 | 3.242 |
| | | | NB2 | 6.917 | 0.429 | 16.136 | <.0001 | 6.212 | 7.622 |
| | | | MT1 | 4.822 | 0.417 | 11.564 | <.0001 | 4.136 | 5.508 |
| | | | MT2 | 2.230 | 0.364 | 6.120 | <.0001 | 1.631 | 2.829 |
| | | | ST | 5.895 | 0.424 | 13.905 | <.0001 | 5.197 | 6.592 |
| | | NB1 | NB2 | 4.296 | 0.434 | 9.909 | <.0001 | 3.583 | 5.009 |
| | | | MT1 | 2.200 | 0.426 | 5.170 | <.0001 | 1.500 | 2.901 |
| | | | ST | 3.273 | 0.431 | 7.586 | <.0001 | 2.563 | 3.983 |
| | | NB2 | MT1 | -2.095 | 0.471 | 0.471 | <.0001 | -2.869 | -1.321 |
| | | | MT2 | -4.687 | 0.427 | -10.990 | <.0001 | -5.389 | -3.986 |
| | | | ST | -1.023 | 0.477 | -2.144 | .03 | -1.807 | -0.238 |
| | | MT1 | MT2 | -2.592 | 0.418 | -6.206 | <.0001 | -3.279 | -1.905 |
| | | | ST | 1.073 | 0.467 | 2.298 | .02 | 0.305 | 1.840 |
| | | MT2 | ST | 3.665 | 0.423 | 8.656 | <.0001 | 2.968 | 4.361 |

*Note: In this table, for the effect of discrete treatment NDRT on the association between **X** and **Y**, the baseline is the control group **T0**.*

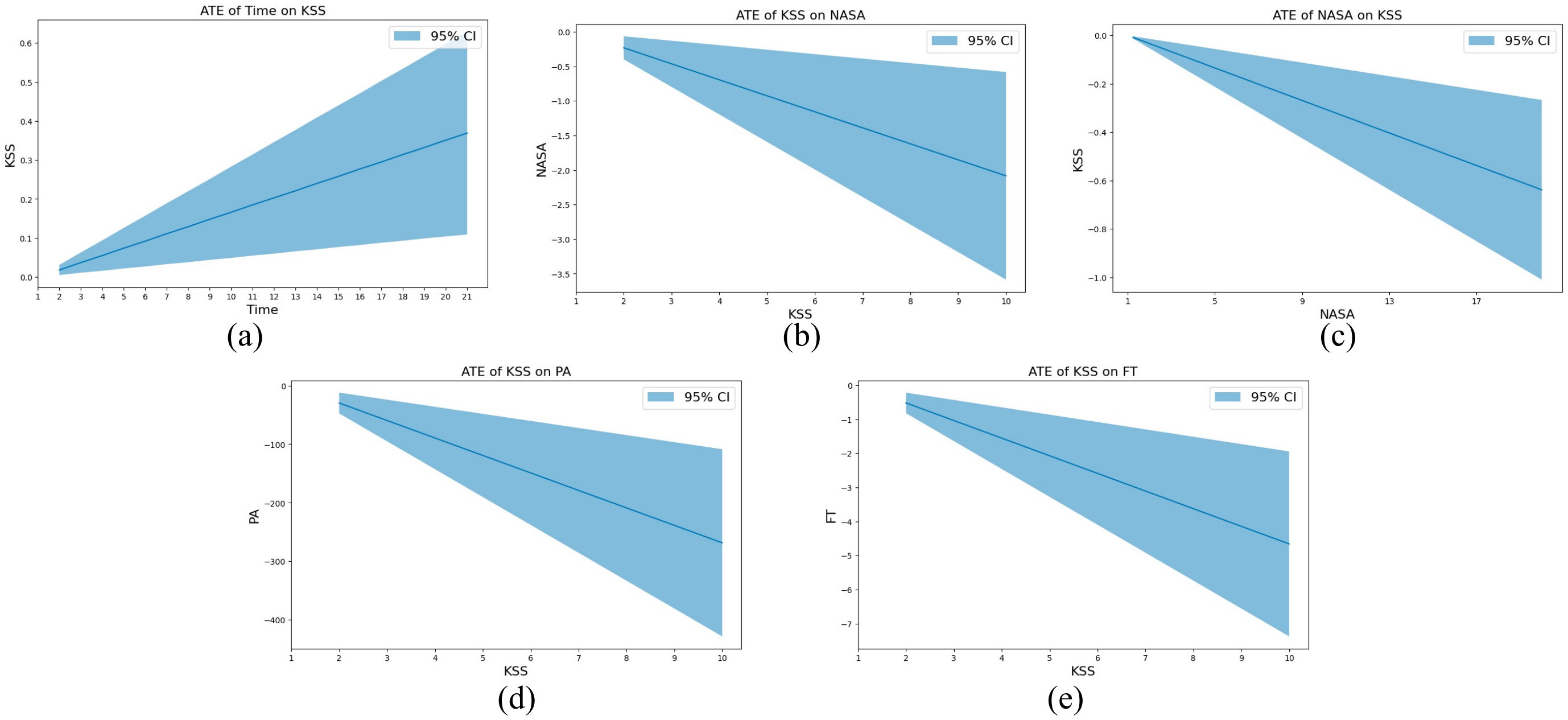

**Fig. 5.** Visualizations of significant ATE of continuous treatments. Subfigure (a) is from Model (a); subfigure (b) is from Model (e); subfigure (c) is from Model (f); and subfigure (d)(e) is from Model (g).

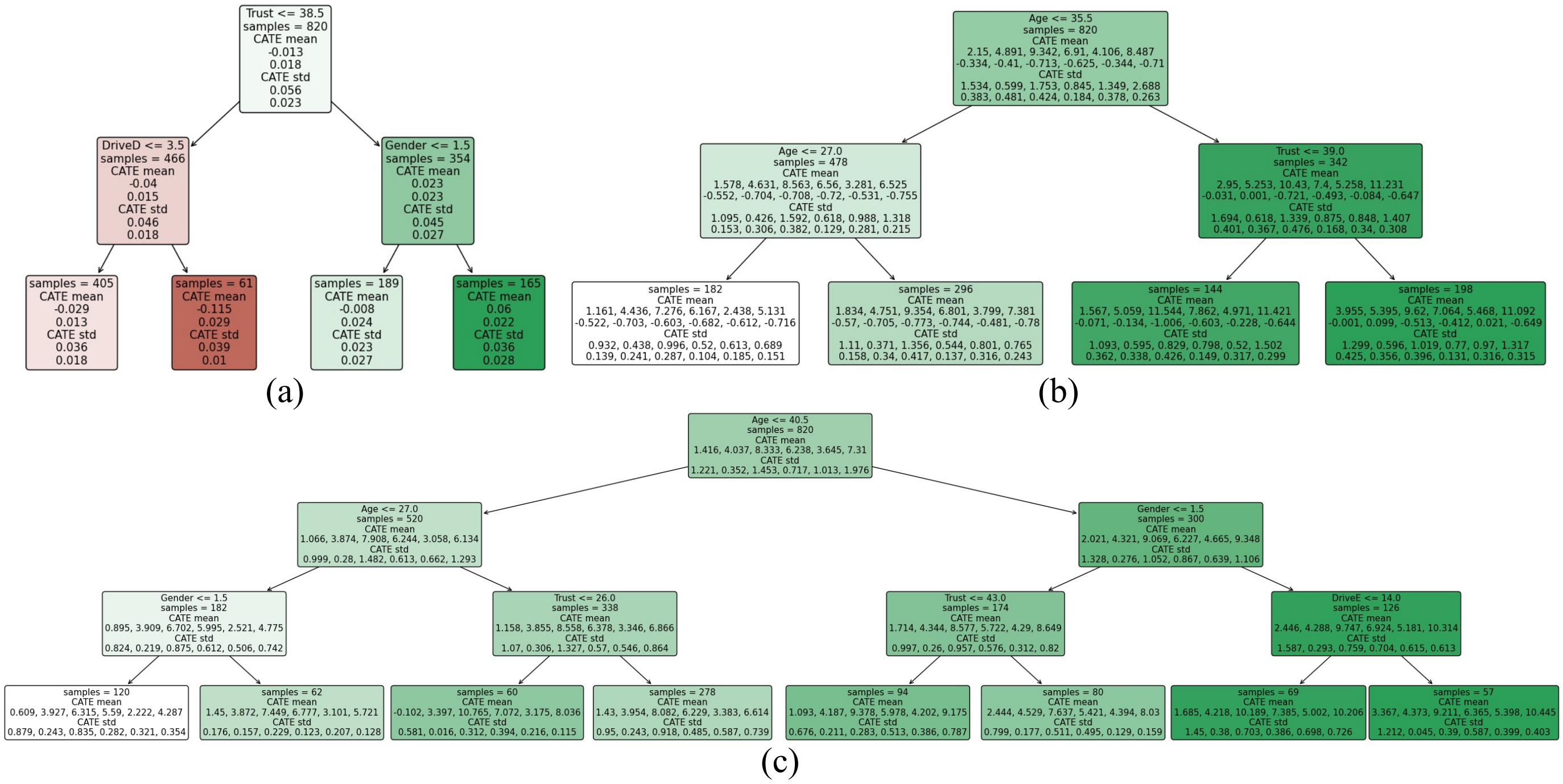

**Fig. 6.** The Heterogeneity tree of CATE over people with different features. In this figure, each node represents a group of samples, and the darker the node color, the greater the change in the outcome (Y) after treatment (T) was applied. The red and green colors represent the overall negative and positive changes in the outcome after a treatment was applied. In each block, all the samples that satisfy the condition in the first line are categorized into the left node; otherwise, they are categorized into the right node. Then, each line under the "CATE mean" and the "CATE std" shows the mean and standard deviation of CATE for each Y, respectively; the numbers in each line correspond to levels in T. Subfigures (a)(b)(c) are based on Model (a)(b)(d) respectively, given no significant influence of T on Y was observed in Model (c). For the specific CATE value, in subfigures (a)(b), the first row of the CATE mean and std value in each node belongs to NASA score, and the second row is KSS. In subfigures (b)(c), each column of the CATE mean and std value corresponds to NB0, NB1, NB2, MT1, MT2, and ST respectively.

## IV. Results

The coefficients between features (X) and the outcome (Y) of each model are presented in Table III. The ATE and CATE are summarized in Table IV and Figure V, respectively. To better illustrate how individual heterogeneity can affect the influence of Time and NDRT on NASA and KSS, we further visualized the CATE over the population with different features in Figure 6.

For example, in response to RQ1, based on the results of Model (a) in Table III, we found that considering the influence of Time on the variation of NASA and KSS, Trust and DriveD were still positively associated with NASA and KSS, respectively. At the same time, a significant effect of Time on KSS was identified (as shown in Figure 5(a)), but not on NASA. Referring to Figure 6(a), we noticed that, for those with high Trust (Trust>38.5), with the increase of Time, their NASA and KSS scores all increased in general. Particularly, those who have low trust and longer driving distances (DriveD), and females with high trust present higher sensitivity to Time (i.e., larger CATE).

Moreover, as shown in Table IV, compared to Base, all NDRTs contributed to higher NASA and lower drowsiness scores. We further visualized the effects of NDRTs in Figure 7. We found that conducting any NDRTs can significantly reduce the KSS score. This effect also varied across tasks: overall, those tasks that led to higher NASA also led to lower KSS, although no significant comparisons exist between groups (e.g., in Figure 7, MT1 led to higher NASA compared to MT2, but did not lead to lower KSS). It is worth noting that as the interactions between KSS and NASA may exist, the accuracy of ATE comparisons between tasks may be inaccurate. Therefore, we built Model (c)(d) and tried to answer RQ2.

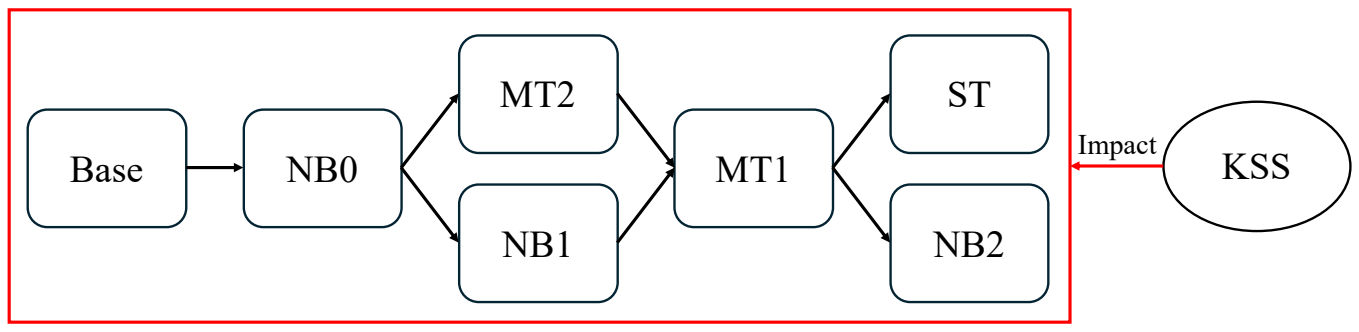

**Fig. 7.** The visualization of the different impacts of NDRTs on NASA. With the black arrow from left to right, the task leads to higher NASA scores than previous tasks. Note that, these effects are still influenced by KSS scores.

According to Table III, when excluding the impact from NASA, the correlation between Individual Characteristics and KSS score in Model (c) no longer existed. Besides, there was no significant ATE of Time on the KSS score. To further verify it, we constructed another model (X= Individual Characteristics; Y=KSS; T=Time; W=NASA and NDRT), and still, no significant ATE of NDRT on KSS was found, which is opposite to the results of Model (b). Other results related to RQ3 and RQ4, can be found from the results of Model (e)-(h) in Table III, IV, and Figure 5. The detailed analysis is included in the Discussion section.

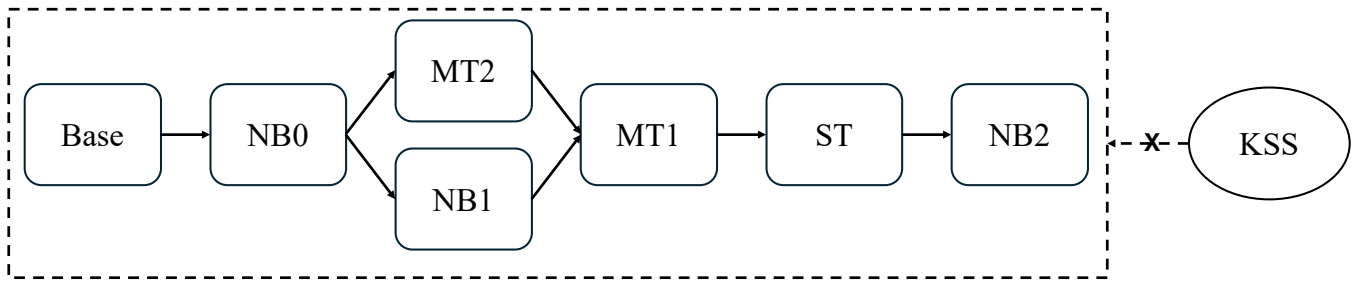

**Fig. 8.** The visualization of the different impacts of NDRTs on NASA without impacts from KSS.

## V. Discussion

*RQ1: How are driver states affected by NDRTs and the progress of the experiment (i.e., time) jointly?*

Through Model (a)(b), we simulated real-world scenarios, where heterogeneous drivers are exposed to multiple surface factors. Firstly, as expected, based on the results of Figure 5(a), we found that drivers were more likely to be drowsy with the increase in driving time. At the same time, as shown in Table IV, NDRTs imposed an additional cognitive load. These findings have been extensively validated by previous studies [9], [10] and can validate the effectiveness of our experiment design. However, our approach has provided more interesting insights on top of the previous research.

Based on the results, we found that the influence of surface factors (i.e., individual characteristics and environmental factors) could be moderated by individual heterogeneity. Referring to Table III and Figure 6 (a)(b), drivers with longer driving mileage were more likely to be drowsy with the increase in driving time, which is different from the conclusion drawn in previous research [29]. However, it should be noted that [29] was conducted in non-automated vehicles. With the assistance of ADAS in our experiment, the relatively low workload experienced by experienced drivers [57] might contribute to their faster development of drowsiness [58], [59]. At the same time, we also observed that drivers with high trust in the ADAS reported higher cognitive load, potentially because they have less concern about ADAS and are more engaged in NDRTs.

At the same time, as shown in Table III and Figure 6, we found that compared to younger drivers, older drivers exhibited higher drowsiness and cognitive load in general when performing NDRTs. The increased self-reported cognitive load is easy to understand, given cognitive capacity decreases with age [27], [28]. However, the drowsiness development was contrary to the findings in previous research in non-automated vehicles and when driving was the sole task (e.g., [9]), where younger drivers were more likely to get drowsy. Again, we suppose that as older drivers have lower cognitive capacity, the NDRTs might have led to cognitive overload, and hence facilitated drowsiness development [25], [26].

*B. Excluding the influence of NDRT and time, whether and how does individual heterogeneity affect cognitive load and drowsiness?*

De-coupling the effect of individual differences on driver states (given that the driving-related and NDRT tasks can affect driver states in addition to individual differences) is difficult in a natural driving environment with the traditional approach but is possible with the DML analysis.

Specifically, excluding the effects of NDRTs and cognitive load (Model (c)), individual heterogeneity, and driving time did not affect drowsiness levels. Further, the effect of NDRTs on drowsiness found in Model (b) disappeared in Model (d), illustrating that in our experiment, the variation in drivers' "drowsiness" might not be attributed to long hours of driving, but extended engagement with NDRTs. This phenomenon indicates that the drowsiness we observed was more likely to be task-related mental fatigue [60] and 1.5 hours of driving might be insufficient to induce non-drowsiness in the SAE L3 vehicles. Future empirical studies should differentiate the type of drowsiness (e.g., mental fatigue and physical fatigue [61]) and further scrutinize the role of different factors on drowsiness development.

Additionally, when the effects from drowsiness and driving time were excluded, we noticed that individual differences (age and trust) and NDRTs were still influential factors of cognitive load. This suggests that individual characteristics affected cognitive workload directly, instead of through indirectly influencing drowsiness levels. Drivers with different capacities likely need to devote different levels of effort to complete the tasks in the experiment.

### *C. Excluding the influence of individual heterogeneity, what are the compound effects of driver states on symbols?*

RQ3 explores the relationships between physiological and eye-tracking measures and drivers' states, which is critical to driver-state monitoring. From Model (e)(f)), we found significant changes in some EDA features (i.e., SCR) when both cognitive workload and drowsiness levels varied (Table III). In addition, we found that cardiac activity exhibited a significant correlation with sleepiness levels in both the value and frequency domains. Specifically, based on Model (f), we found that given a cognitive workload (i.e., NASA value), the higher drowsiness levels (i.e., KSS) were correlated with lower SCR, higher RMSSD and HF, and lower SDNN, which were also widely used for fatigue and cognitive load monitoring [41], [42]. These findings were in line with previous research [62], [63]. Besides, through Model (e)(f), we further validated the overall negative correlation that exists between drowsiness and cognitive workload based on Figure 5 (b)(c).

### *D. Excluding the influence of individual heterogeneity and cross-effect between states, what are the relationships between symbols and a single state?*

To pinpoint the relationship between certain driver states and symbols, we first isolated the effect of drowsiness when exploring the effect of cognitive workload and found no significant symbols. This is contradictory to some previous studies that have observed the association between cognitive load and physiological metrics, such as pupil diameter [8]. Likely, the effect of cognitive load observed in previous studies was actually the joint effect of cognitive load and drowsiness. Future experiment design is needed to disentangle the two states and scrutinize the effects. At the same time, conversely, after eliminating the effect of cognitive workload, we found some eye-movement features were significantly correlated to drowsiness levels. This finding agrees with previous research which found that the increased drowsiness levels were associated with changes in attention allocation strategies [45]. Thus, from the driver state estimation perspective of view, our findings indicate that the overall drowsiness (which can be moderated by cognitive workload) may be effectively estimated by EDA and cardiac-related metrics (see Table III). Whereas the influence of non-cognitive-related drowsiness (i.e., drowsiness excluding cognitive load) should better be measured by eye-tracking metrics.

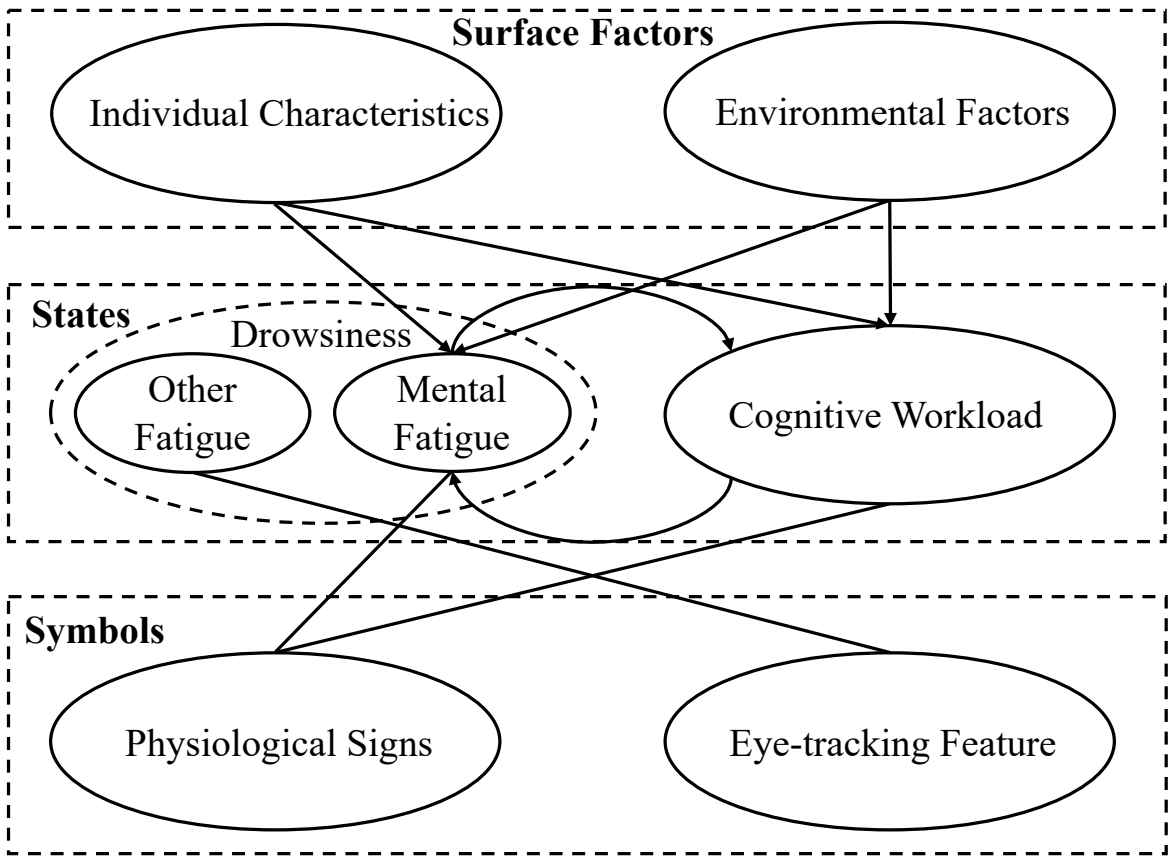

**Fig. 9.** The updated causality relation graph based on findings in this work.

## VI. LIMITATIONS

It is worth noting that this paper has some limitations. Firstly there was still a gap in participants' perception and performance in the driving simulator and the real driving environment [63]. Further, considering that the participants in this study were all from mainland China, and taking into account the potential regional and ethnic effects on drivers' behaviors and states, future experiments with a wider range of people in a real road environment should be conducted to further validate the conclusions from this paper.

## VII. CONCLUSION

In a real-world driving environment, multiple states often occur simultaneously, subject to individual differences and environmental factors. Although previous studies have noted a correlation between drowsiness and cognitive load, due to limitations in experimental design and analytical methods, the effects reported in previous studies focusing on a single state may be a joint effect from multiple states. To address these problems, this paper first introduced the DML analysis method in the field of driver state analysis. Based on an L3 autonomous driving simulator experiment with 48 participants and eight DML models, we first analyzed the individual and environmental factors (including multidimensional NDRTs ansd driving time) that may affect the variation of driver states. Subsequently, by setting confounding factors, we isolated the influence of individual and environmental factors on drowsiness and cognitive workload, when the effects of other states were eliminated. Further, the complex causal relationship between drowsiness and cognitive load was successfully decoupled. In addition, we investigated key physiological and eye-tracking indicators in the presence of cross-effects between states, as well as under the influence of a single state (see Figure 9). In general, our findings empirically demonstrated the co-occurrence of multiple states in drivers and explored the causal relationship between multiple states and driver physiological and eye-tracking features. The causal inference analytical framework introduced in this paper also provides insights for subsequent analytical work. At the same time, the state-related metrics identified in this paper can facilitate the development of more fine-grained driver-state monitoring systems.

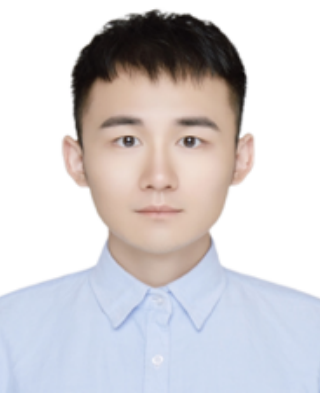

**Jiyao Wang** received the B.Eng. degree in Software Engineering from the Sichuan University, Chengdu, China in 2021, and M.Sc. degree in Big Data Technology from the Hong Kong University of Science and Technology, Hong Kong S.A.R., China, in 2022. Currently, he is a Ph.D. student in the Robotics and Autonomous Systems Thrust at the Hong Kong University of Science and Technology (Guangzhou).

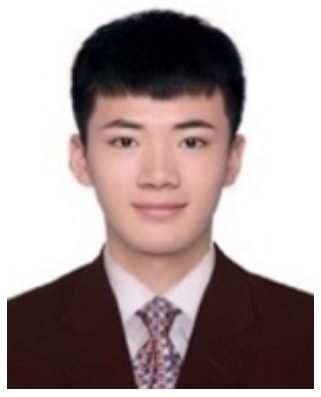

**Ange Wang** received his bachelor's degree from East China Jiaotong University in 2019 and his M.Sc. degree from the Beijing University of Technology in 2022. He is now a Ph.D. student in the Intelligent Transportation Thrust at the HKUST (GZ).

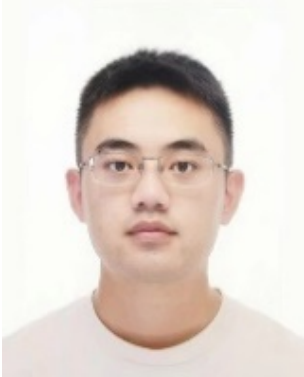

**Song Yan** received his bachelor's degree in automotive engineering from Zhejiang University, China, and master's degree in mechanical engineering from the University of Tokyo, Japan. He is currently a PhD student in Robotics and Autonomous Systems Thrust at the HKUST (GZ).

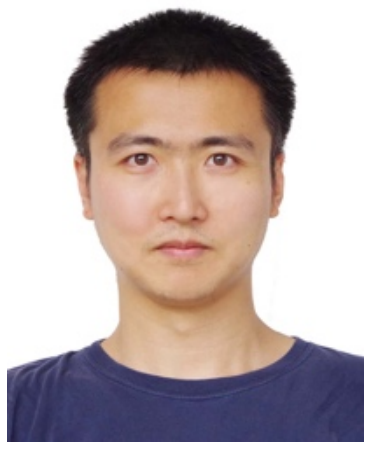

**Dengbo He** received his bachelor's degree from Hunan University in 2012, M.S. degree from Shanghai Jiao Tong University in 2016, and Ph.D. degree from the University of Toronto in 2020. He is currently an assistant professor from the Intelligent Transpiration Trust and Robotics and Autonomous Systems Thrust, the HKUST (GZ). He is also affiliated with the Department of Civil and Environmental Engineering, HKUST, Hong Kong SAR.

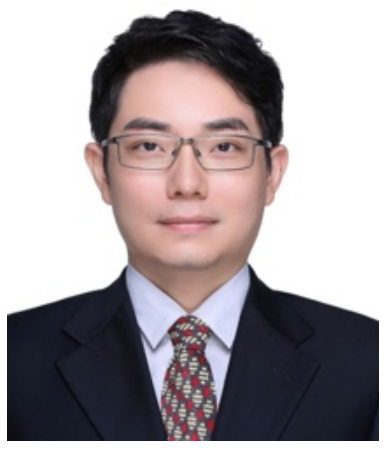

**Kaishun Wu** (Fellow, IEEE) received the Ph.D. degree in computer science and engineering from Hong Kong University of Science and Technology (HKUST), Hong Kong, in 2011. He was a Distinguished Professor and the Director of Guangdong Provincial Wireless Big Data and Future Network Engineering Center with Shenzhen University, Shenzhen, China. In 2022, he joined HKUST (GZ) as a Full Professor with DSA Thrust and IoT Thrust. He is an IET, AAIA, and IEEE Fellow.